\documentclass{emulateapj}
\usepackage{epsfig}
\usepackage{color}
\usepackage{xcolor}

\shorttitle{HCN and SFR in M51}
\shortauthors{Chen et al.}

\begin{document}
\nocite{*}   

\title{Dense Gas in the Outer Spiral Arm of M51}

\author{Hao Chen\altaffilmark{1, 2, 3,*}}

\author{Jonathan Braine\altaffilmark{4, 5}}

\author{Yu Gao\altaffilmark{6, 7}}


\author{Jin Koda\altaffilmark{8}}

\author{Qiusheng Gu\altaffilmark{1, 2, 3,*}}

\email{* chenhao198701@yahoo.com, qsgu@nju.edu.cn}

\altaffiltext{1}{School of Astronomy and Space Science, Nanjing University, Nanjing 210093, P. R. China}
\altaffiltext{2}{Key Laboratory of Modern Astronomy and Astrophysics, Nanjing University, Nanjing 210093, P. R. China}
\altaffiltext{3}{Collaborative Innovation Center of Modern Astronomy and Space Exploration， Nanjing, 210093, P.R. China}
\altaffiltext{4}{Univ. Bordeaux, Laboratoire d'Astrophysique de Bordeaux, F-33270, Floirac, France.}
\altaffiltext{5}{CNRS, LAB, UMR 5804, F-33270, Floirac, France}
\altaffiltext{6}{Purple Mountain Observatory, Chinese Academy of Sciences, 2 West Beijing Road, Nanjing 210008, P. R. China}
\altaffiltext{7}{Key Laboratory of Radio Astronomy, Chinese Academy of Sciences, Nanjing 210008, P. R. China}
\altaffiltext{8}{Department of Physics and Astronomy, Stony Brook University, Stony Brook, NY 11794-3800}


\begin{abstract}
There is a linear relation  between the mass of dense gas, traced by the HCN(1--0) luminosity, and the star formation rate (SFR), traced by the far-infrared luminosity. Recent observations of galactic disks have shown some systematic variations. 
In order to explore the SFR--dense gas link at high resolution ($\sim 4$\arcsec, $\sim 150$ pc) in the outer disk of an external galaxy, we have mapped a region about 5 kpc from the center along the northern spiral arm of M51 in the HCN(1--0), HCO$^+$(1--0) and HNC(1--0) emission lines using the Northern Extended Millimeter Array (NOEMA) interferometer. The HCN and HCO$^+$ lines were detected in 6 giant molecular associations (GMAs) while HNC emission was only detected in the two brightest GMAs. 
One of the GMAs hosts a powerful H\uppercase\expandafter{\romannumeral2} region and HCN is stronger than HCO$^+$ there. 
Comparing with observations of GMAs in the disks of M31 and M33 at similar angular resolution ($\sim 100$  pc), we find that GMAs in the outer disk of M51 are brighter in both HCN and HCO$^+$ lines by a factor of 3 on average.
However, the $I_{\rm HCN}/I_{\rm CO}$ and $I_{\rm HCO^+}/I_{\rm CO}$ ratios are similar to the ratios in nearby galactic disks and the Galactic plane. 
Using the Herschel 70 $\mu$m data to trace the total IR luminosity at the resolution of the GMAs, we find that both the L$_{\rm IR}$--L$_{\rm HCN}$ and L$_{\rm IR}$--L$_{\rm HCO^+}$ relations in the outer disk GMAs are consistent with the proportionality between the L$_{\rm IR}$ and the dense gas mass established globally in galaxies within the scatter. The IR/HCN and  IR/HCO$^+$  ratios of the GMAs vary by a factor of 3, probably depending on whether massive stars are forming or not.

\end{abstract}


\section{INTRODUCTION}

Stars form in the dense cores of giant molecular clouds (GMCs).  The dense cores are traced by high dipole-moment molecules like HCN, HCO$^+$, and CS \citep{Evans1999ARA&A..37..311E}.  In cold regions where the density is very high, a commonly used probe in the Galaxy is N$_2$H$^+$, because it does not deplete onto dust grains.  In the warm regions surrounding massive protostars, high-$J$  CO lines are useful probes of the density and temperature \citep[in the sub-millimeter regime,][]{Lu2014ApJ...787L..23L, Liudz2015ApJ...810L..14L, zhao2016ApJ...820..118Z}.
The dense gas tracer HCN (n$_{\rm eff} \sim 10^5 cm^{-3}$) exhibits the strongest line in galaxies after CO and $^{13}$CO. In intensely star-forming galaxies such as ultra-luminous infrared galaxies (ULIRGs) the HCN line can be stronger than the $^{13}$CO \citep[e.g.,][]{Baan2008A&A...477..747B}.  A linear relationship between the SFR, traced by infrared luminosity, and dense gas mass, traced by the HCN luminosity is found in galaxies \citep{Gao2004ApJ...606..271G, Zhang2014ApJ...784L..31Z, Liulj2015ApJ...805...31L} and Galactic clumps \citep{Wu2005ApJ...635L.173W, Wu2010ApJS..188..313W}. 
The IR-to-HCN(1--0) luminosity ratio is a proxy for the ratio of the star formation rate (SFR) and dense gas mass, referred to as the star formation efficiency of dense molecular gas, SFE$_{dense}$, and this ratio is almost constant in galaxies. 
The strong linear relation between HCN intensity and IR emission, even in ULIRGs where the IR-CO relation becomes non-linear, suggests that it is the mass of dense gas rather than the molecular gas reservoir \citep{Solomon1992ApJ...387L..55S, Gao2007ApJ...660L..93G} that governs star formation.

From large-scale mapping of the HCN emission in M51, \citet{Chen2015ApJ...810..140C} and \citet{Bigiel2016ApJ...822L..26B}
showed that the IR-to-HCN ratio (SFE$_{dense}$) is lower in the central kpc than the outer disk. The HCN emission is also strong compared to CO in the central kpc of M51.  
If the SFE$_{dense}$ is in fact {\it not} constant, then either HCN(1--0) is not a reliable measure of the dense gas mass throughout the disk or other factors \citep[e.g. turbulence enhanced by the shear motion,][]{Krumholz 2015MNRAS.453..739K} prevent dense gas from turning into stars.
This effect is not limited to M51.
The Antennae galaxies (NGC 4038/4039) show that the IR-to-HCN ratio is on average 4 times higher in the 3 overlap regions than in the two nuclei \citep{Bigiel2015ApJ...815..103B}.
For 48 HCN detections of 29 nearby disk galaxies in the HERACLES survey, \citet{Usero2015AJ....150..115U} found that the IR-to-HCN ratios increase systematically with radius in galaxies, 6--8 times lower near galaxy centers than in the disks.
\citet{Longmore2013MNRAS.429..987L} stated that the current SFR in the inner 500 pc of our Galaxy is ten times lower than the rate predicted from the dense gas.

\begin{figure*}
\centering\includegraphics[angle=0,scale=.9]{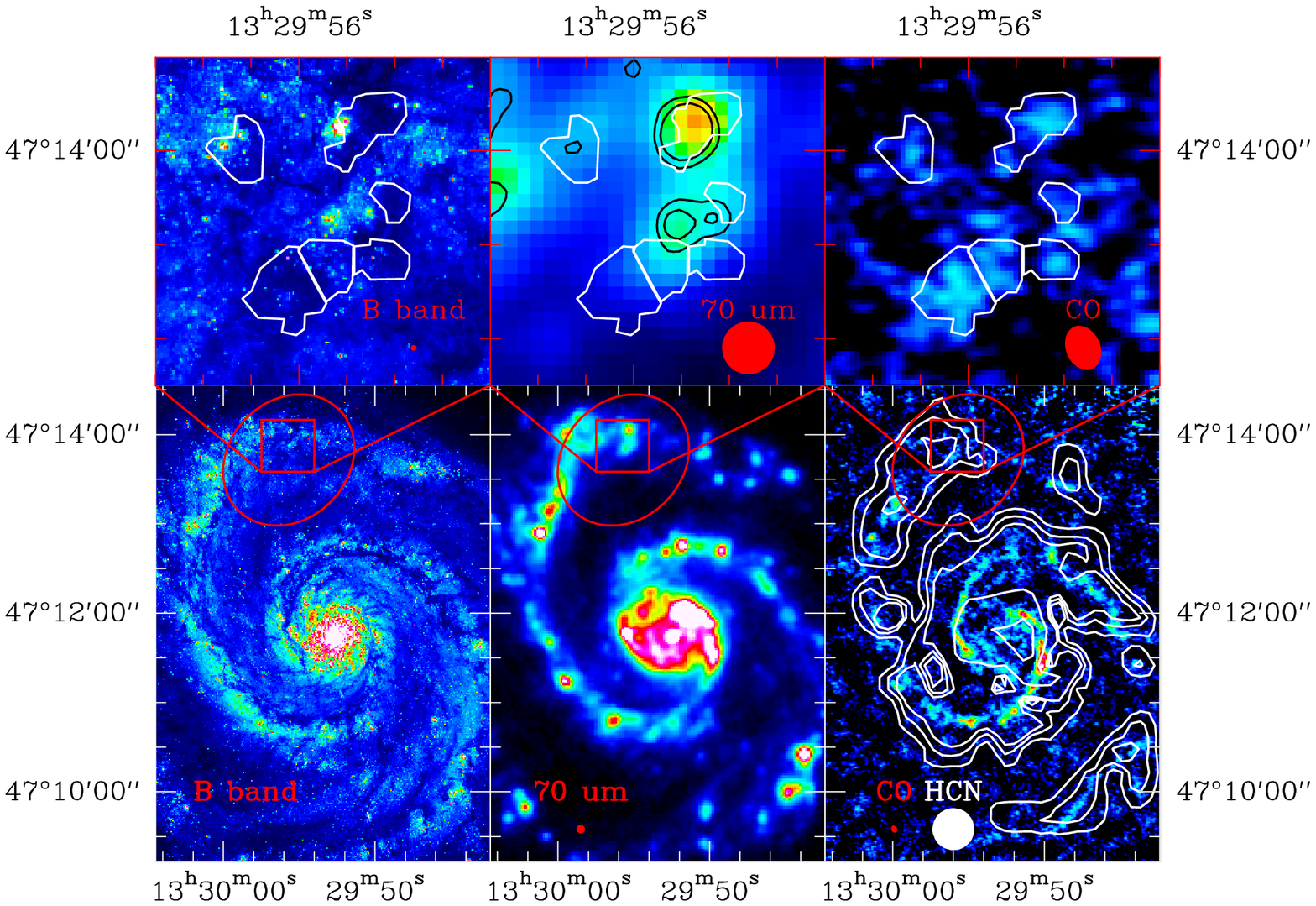}
\caption{The bottom panels are {\it left} B band (4350\AA) image from the Hubble Space Telescope \citep{Mutchler2005AAS...206.1307M}, {\it middle} 70 $\micron$ image from the Herschel telescope and {\it right} CO 1--0 image from CARMA array \citep{Koda2011ApJS..193...19K} with HCN contours from IRAM 30M telescope shown \citep{Chen2015ApJ...810..140C}.
The area we observed is indicated by red ellipses.
The upper panels are zooms of the area in the red square. 
The white contours in the upper panels show the GMAs defined in Figure 2.
The zoom of 70 $\micron$ image is shown with black contours representing the H$\alpha$ intensity \citep{Rand1992AJ....103..815R}.
The beam size of each map is shown at the the bottom of each image.}
\end{figure*}

The IR--HCN relation in the outer disk at high resolution is not well known because the HCN emission in the outer disk is much weaker than the center \citep{Curran2001A&A...368..824C, Gao2004ApJS..152...63G, Chen2015ApJ...810..140C} and there are few high resolution dense gas observations towards the outer disk except for the very nearby galaxies, M31 \citep{Brouillet2005A&A...429..153B} and M33 \citep{Rosolowsky2011MNRAS.415.1977R, Buchbender2013A&A...549A..17B}. 
In this work, we present observations at 150 pc resolution of a region along the northern spiral arm in the outer disk ($\sim$ 5 kpc from the center) of M51 in HCN(1--0), HCO$^+$(1--0), and HNC(1--0), all of which trace dense molecular gas.  
HCN and HNC are isomers, and have similar energy spectra and dipole moments. 
HCO$^+$ has a slightly lower dipole moment and, although it is an ion, appears to also trace dense gas quite well \citep{Jiang2011MNRAS.418.1753J}. With these observations, we can fill the gap (between GMC and kpc scale region in galactic disk, $10^7-10^8 L_\odot$ in IR) in IR--HCN and IR--HCO$^+$ relations.

This region was selected because it is in the outer disk and for its high signal-to-noise HCN spectrum observed with IRAM 30M (Chen et al 2015).  The metallicity is solar to within the uncertainties \citep{Bresolin2004ApJ...615..228B}.
With these data in a rather small region (in M51), we obtain several independent data points without the potential influence on star formation by the radial (or other) variations of turbulence, metallicity and pressure \citep{Usero2015AJ....150..115U, Chen2015ApJ...810..140C, Bigiel2016ApJ...822L..26B}.

\begin{figure*}
\begin{center}
\includegraphics[width=0.8\textwidth]{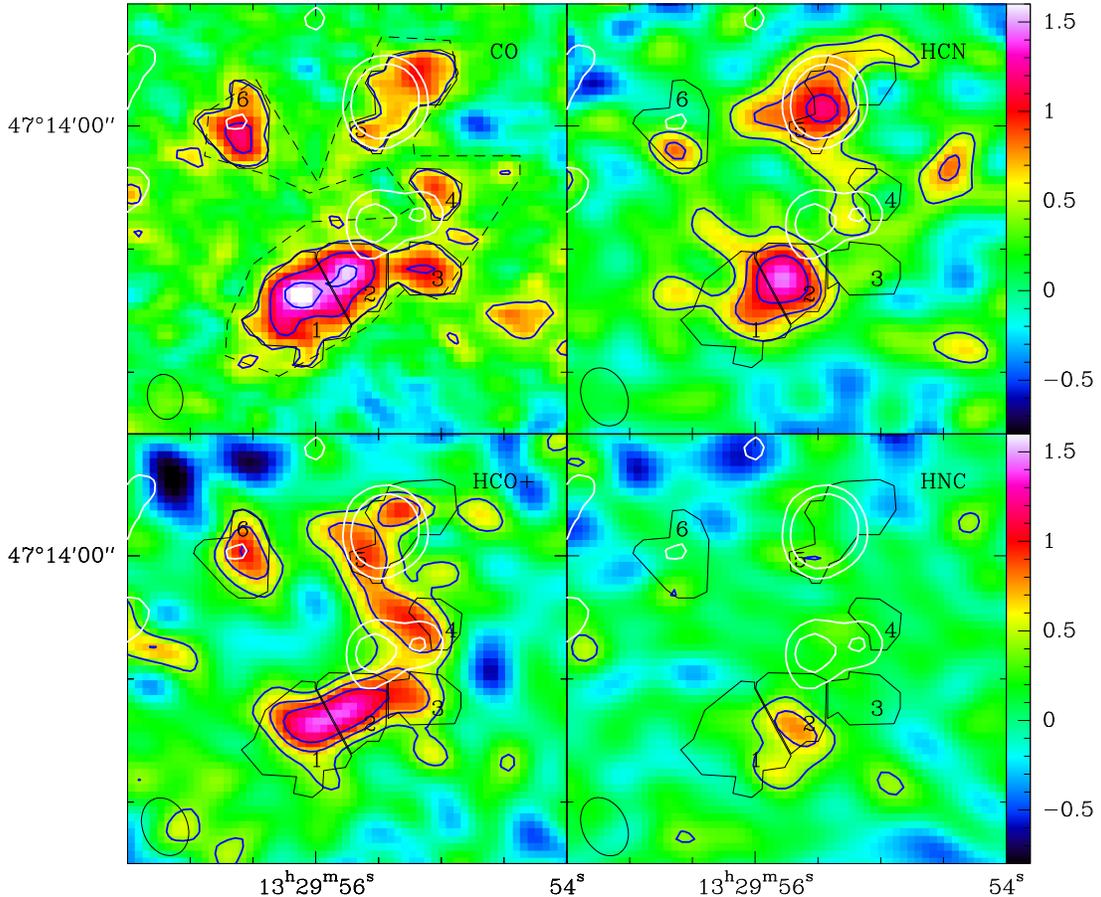}
\caption{CO, HCN, HCO$^+$ and HNC integrated intensity maps. All the maps are in the units of K\ km/s and show the restoring beam size to the lower left of each panel and the line name to the upper right. 
The color wedge for the HCN, HCO$^+$ and HNC maps is indicated to the right (-0.8 to 1.6 K\ km/s). 
The CO color wedge goes from -40 to 80 K\ km/s with the same linear color coding.  
3, 5 and 7 $\sigma$ contours are shown in blue for the HCN, HCO$^+$ and HNC images, and the CO image shows the 5 ,9 and 13 $\sigma$ contours. 
The black dashed contour in the CO map shows the polygon, defined from the CO image, used for CLEANing the HCN, HCO$^+$ and HNC data. 6 black numbered polygons show the GMAs we use for the calculations and to show the spectra.
The same H$\alpha$ contours as in Figure 1 are shown in white over all the images in order to compare sites of star formation with the molecular gas tracers.
} 
\end{center}
\end{figure*}

\begin{deluxetable}{cccccc}
\tabletypesize{\scriptsize}
\tablecaption{CO, HCN, HCO$^+$ and HNC observations}
\tablewidth{0pt}
\tablehead{
\colhead{Line}  & \colhead{Velocity} & \colhead{Velocity} &  \colhead{Beam} &  \colhead{data cube} \\
\colhead{Name}   & \colhead{Resolution} & \colhead{Range} &  \colhead{Size} &  \colhead{rms} \\ 
\colhead{}   & \colhead{[km/s]} &  \colhead{[km/s]} &  \colhead{} &  \colhead{[mK]} 
}
\startdata
\colhead{CO}  & 5.1 & 380.6--420.4 & 3.68\arcsec $\times$ 2.87\arcsec    & 395.3 \\\colhead{HCN}  & 4.2 & 385.4--421.5 & 4.88\arcsec $\times$ 3.67\arcsec  & 12.0 \\
\colhead{HCO$^+$}    & 4.2  & 386.9--419.8 & 4.85\arcsec $\times$ 3.64\arcsec & 11.7 \\\colhead{HNC}  & 6.6  & 394.2--412.6 & 4.96\arcsec $\times$ 3.59\arcsec  &  11.3 
\enddata
\tablecomments{The CO data were obtained by \citet{Koda2011ApJS..193...19K}. The Jy/K conversion factors are 8.71, 8.69, 8.70, and 8.35 for the CO, HCN, HCO$^+$, and HNC data cubes, respectively.}
\end{deluxetable}

\section{OBSERVATIONS AND DATA REDUCTION}

\subsection{HCN(1--0), HCO$^+$(1--0), HNC(1--0)}

The two field mosaic was observed with the NOEMA interferometer in the C and D configurations during 8 short (1--4 hour long) sessions between July 2014 and April 2015.  The size and shape of the mosaic are shown as an ellipse in Figure 1.  The fields were placed at  (8\arcsec, -8\arcsec) and (-8\arcsec, 8\arcsec) offset from 13:29:55.7, 47:13:43.0 (J2000.0).  The standard line and WIDEX correlators were used with spectral resolution of 1.25 and 2 MHz, respectively. The only lines detected were HCN(1--0), HCO$^+$(1--0) and HNC(1--0).

The calibration of the uv data was done using the GILDAS\setcounter{footnote}{0}\footnote{http://www.iram.fr/IRAMFR/GILDAS} software package CLIC and the standard pipeline. 
J1259+516 and/or 1418+546 served as phase and amplitude calibrators and were observed every 20 minutes between source observations.
The flux of 1418+546 was 10\% lower for the data observed on October 17, 2014 than for the other days, so we used J1259+516 as calibrator for the October 17 data.
The uncertainty of the absolute flux calibration is about 10\% at 3 mm.

For imaging and cleaning we used the GILDAS software package MAPPING. The two pointings are combined together to create the dirty map including the primary beam correction. Natural weighting was used to obtain the best signal to noise ratio. After imaging, we ran the CLEAN algorithm using the HOGBOM method on the central part of the field for 10 iterations.
Then, for the channels where flux was detected, we cleaned carefully using the CO map to guide the CLEAN algorithm.  The region CLEANed is shown as a black dashed line in the CO panel of Fig. 2, corresponding roughly to the 5 sigma contour of the CO integrated intensity map.  The results for each channel were checked to make sure that the algorithm gave proper results. 
We chose the best result by comparing with the dirty beam, noise levels and adjacent channels to identify whether further cleaning was necessary.
The noise level is 10\% better when we cleaned with more iterations ($\sim$2000, reaching the default threshold) than with few iterations ($\sim$100) and the fluxes are stable. 
The data presented here represent what we estimate as the most reliable reduction but certainly underestimating slightly the true flux density for the HCN, HCO$^+$ and HNC lines because some residual sidelobes are still present. 
The Jy/K conversion factors are 8.71, 8.69, 8.70, and 8.35 for the CO, HCN, HCO$^+$, and HNC data cubes, respectively.
The spectral resolution, beam size, and $rms$ of the cleaned data cubes are shown in Table 1.

The HCN flux with the IRAM 30M telescope centered at 13:29:66.64, 47:13:58.0 (J2000.0) is 0.48 K km/s, or 2.3 Jy km/s.  The flux of the NOEMA observation is 1.1 Jy K km/s in the same region.  The difference could be due to missing short spacings, such that larger structures are resolved out, but could also be due to the residual interferometric ``bowl", which has not been completely eliminated by the cleaning.

\subsection{CO and IR Data} 

The CO J = 1 - 0  data were taken with the CARMA array combined with zero-spacings from the $5\times5$ Beam Array Receiver System on the Nobeyama Radio Observatory 45M telescope \citep{Koda2011ApJS..193...19K}.
\citet{Schinnerer2013ApJ...779...42S} observed M51 with the Plateau de Bure interferometer at high resolution but their map does not extend to regions as far out as this one.

To compare with the sites of recent star formation, we use the 70 $\micron$ image from the Very Nearby Galaxy Survey (VNGS) which was accessed through the Herschel Database in Marseille (HeDaM\footnote{http://hedam.lam.fr}$^,$ \footnote{The Herschel Database in Marseille (HeDaM) is operated by CeSAM and hosted by the Laboratoire d'Astrophysique de Marseille. }). The resolution is only slightly poorer than ours, so that we can use the Herschel 70$\mu$m images to estimate the star formation rates \citep{Boquien2011AJ....142..111B, Galametz2013MNRAS.431.1956G}.

\begin{figure}
\centering\includegraphics[angle=0,scale=.7]{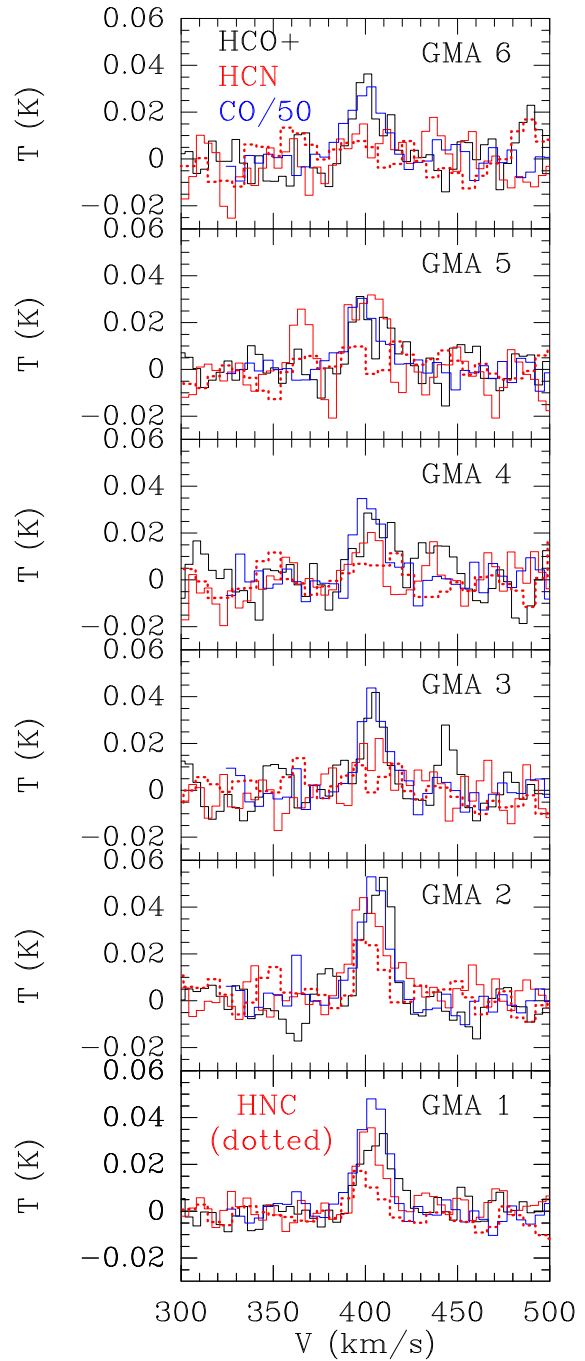}
\caption{CO (blue), HCN (red), HCO$^+$ (black) and HNC (dotted red) spectra for each GMA. 
For calculating the spectra, the CO data cube is smoothed to $4.9\arcsec\times3.6\arcsec$ to match the angular resolution of HCN, HCO$^+$ and HNC data cubes.}
\end{figure}

\section{RESULTS}

Figure 2 shows the integrated intensity maps of the HCN, HCO$^+$ and HNC emission lines along with the CO(1--0) image obtained by \citet{Koda2011ApJS..193...19K}. The integrated intensities were measured as $I = \int Td$V and the emission velocity ranges (V) were defined from the data cube and shown in Table 1.
Uncertainties are calculated as $\delta =  T_{\rm rms}\sqrt{W\delta_c}$ ($T_{\rm rms}$ is the data cube rms, $W$ and  $\delta_c$ are the velocity range and velocity resolution of the integrated intensity map)  and are 5.6, 0.15, 0.14 and 0.13 K km/s for CO, HCN, HCO$^+$ and HNC, respectively.  
Because the HCN, HCO$^+$ and HNC emissions show similar but clearly different morphologies, we chose not to bias our results towards one or another of these tracers but rather use the higher S/N CO map with slightly better angular resolution to define the giant molecular associations (GMAs). 
A total of 6 GMAs have been identified (black numbered polygons in Fig. 2) by the ClumpFind\footnote{http://www.ifa.hawaii.edu/users/jpw/clumpfind.shtml} algorithm \citep{Williams1994ApJ...428..693W}.
The HCN and HCO$^+$ emission peaks are similar at about 1.2 and 1.4 $K\ km/s$ and the peak positions are consistent with each other at the center of GMA 2.  The HNC line is only detected in GMA 1 and 2, and the emission region is displaced to the south compared to the other lines.  GMA 3, 4 and 6 are weak in HCN but strong in HCO$^+$ (and CO). The HCO$^+$ distribution is generally broader than that of the CO, HCN and HNC. 
Figure 2 also shows the H$\alpha$ emission regions (white contours) in order to allow a comparison with the sites of massive star formation.  As can be seen from the zoom in Fig. 1, the 70 $\mu$m emission distribution is quite similar to that of H$\alpha$, so the white contours provide a good reference for the sites of star formation in these GMAs.
GMA 5 clearly shows strong star formation, and its HCN emission is stronger than HCO$^+$ and more centered on the H$\alpha$ than CO.
This may point towards HCN as being more linked to star formation than other dense gas tracers, at least in high ($\sim$solar) metallicity environments.  Clearly more high resolution studies are required to clarify this.  

Spectra for each line integrated over the area of each GMA are shown in Figure 3 and the integrated intensities and uncertainties are provided in Table 2.
As can be seen, the lines are as expected all at the same velocity and with similar line widths. Because several clouds are probably included within one GMA, the hyperfine structure of the HCN line is not visible and the broadening (due to the presence of three components) is not detectable.

\section{ANALYSIS AND DISCUSSION}

\begin{figure}
\centering\includegraphics[angle=0,scale=.35]{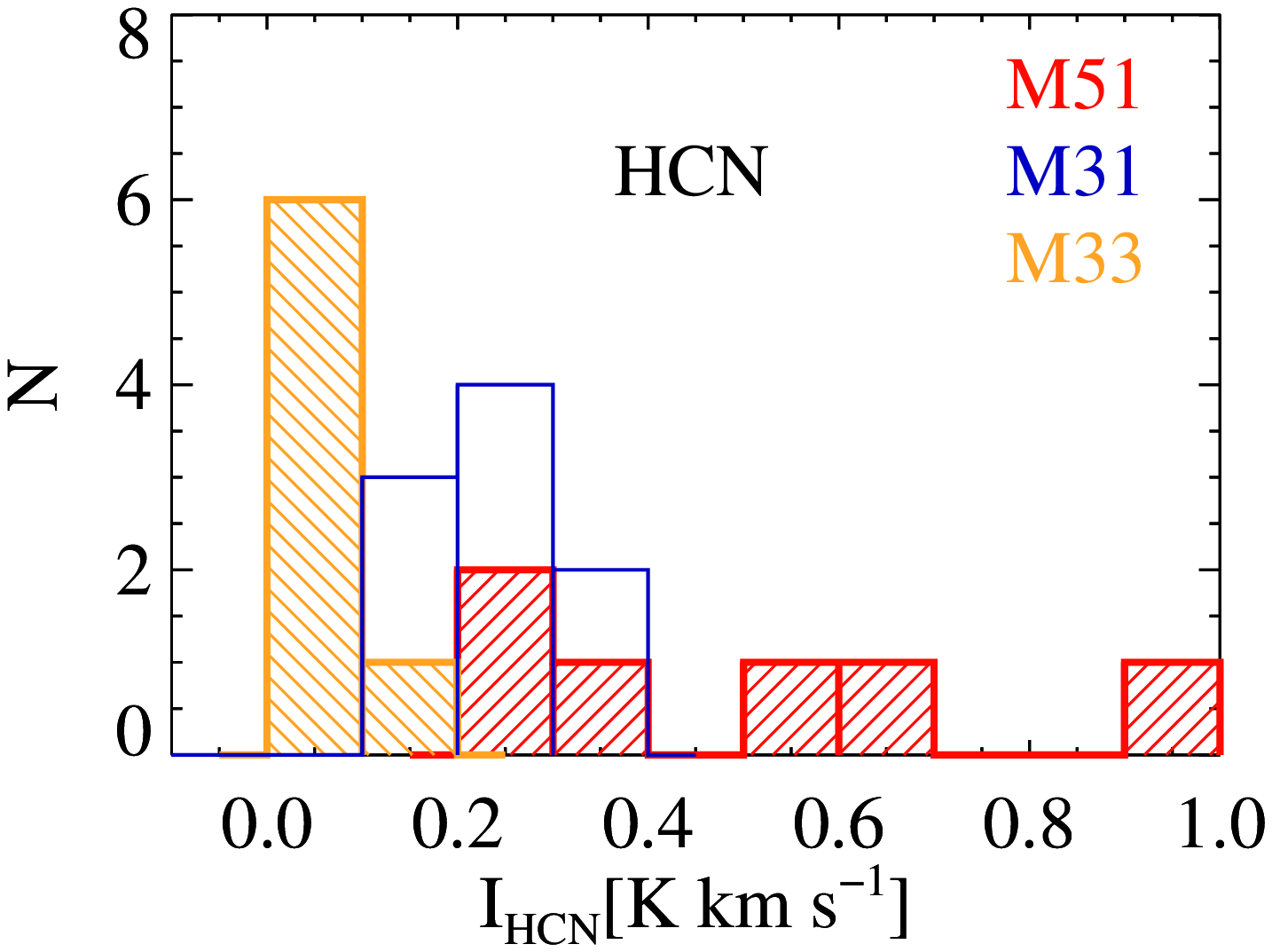}
\centering\includegraphics[angle=0,scale=.35]{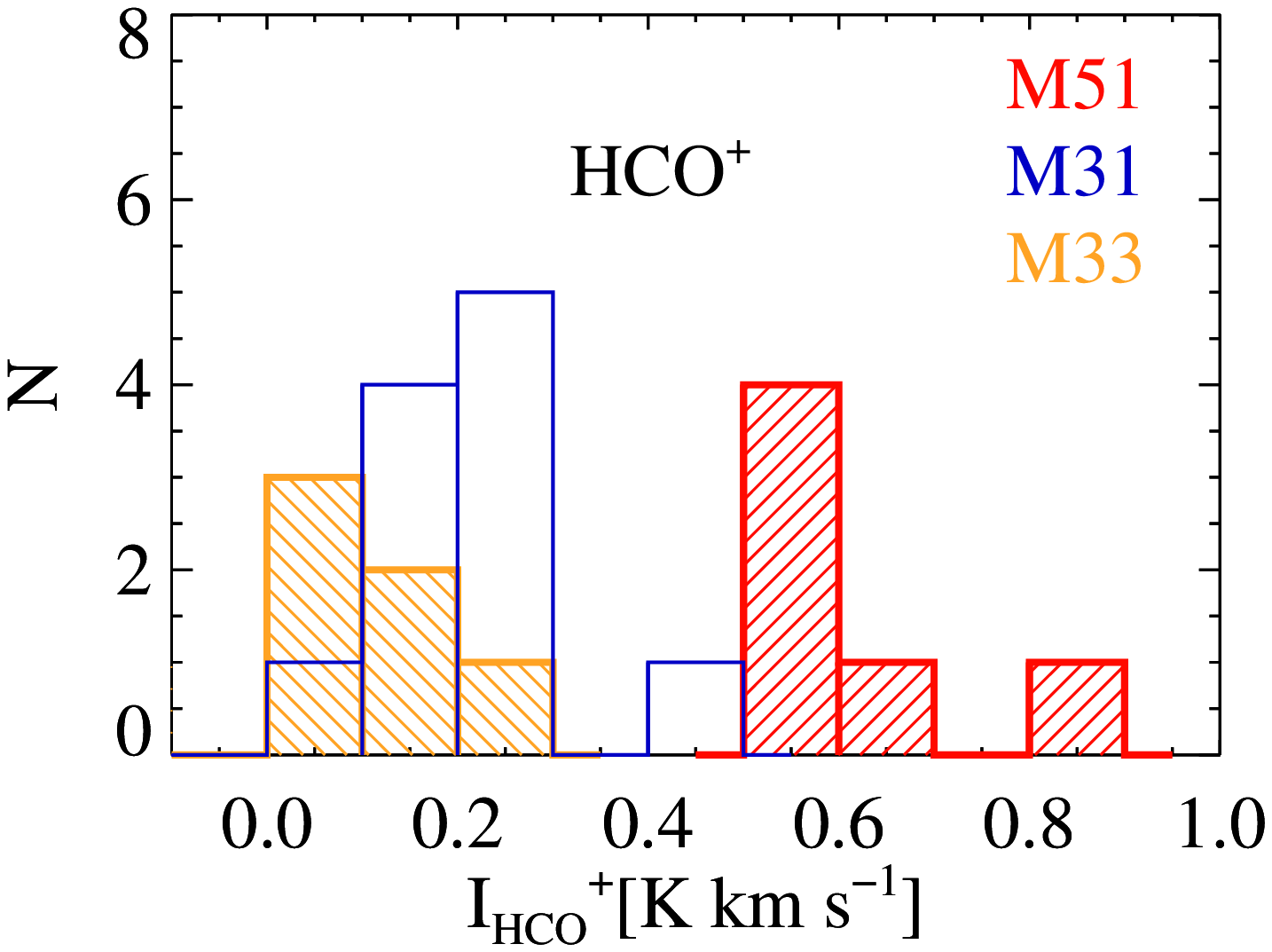}
\centering\includegraphics[angle=0,scale=.35]{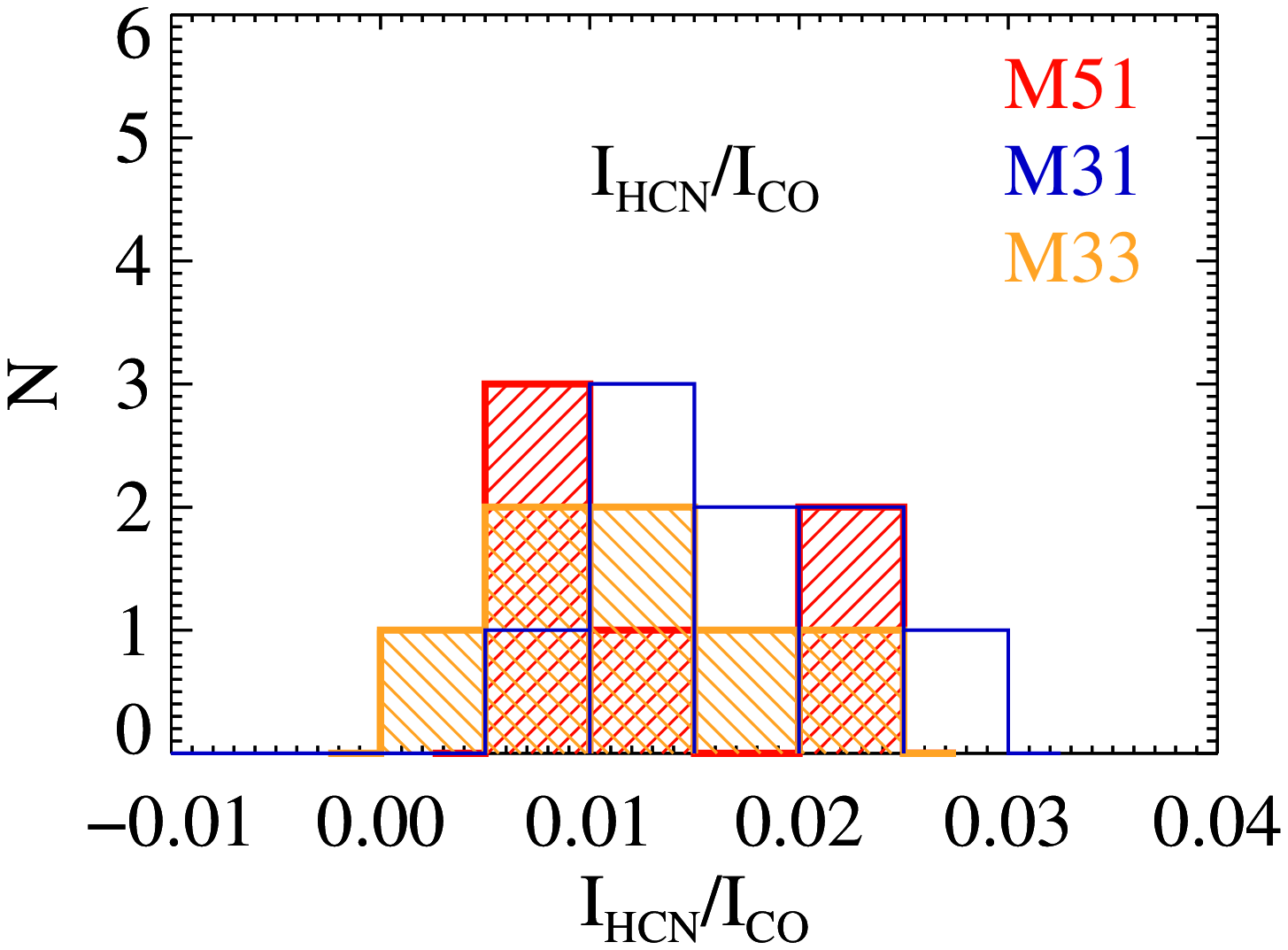}
\centering\includegraphics[angle=0,scale=.35]{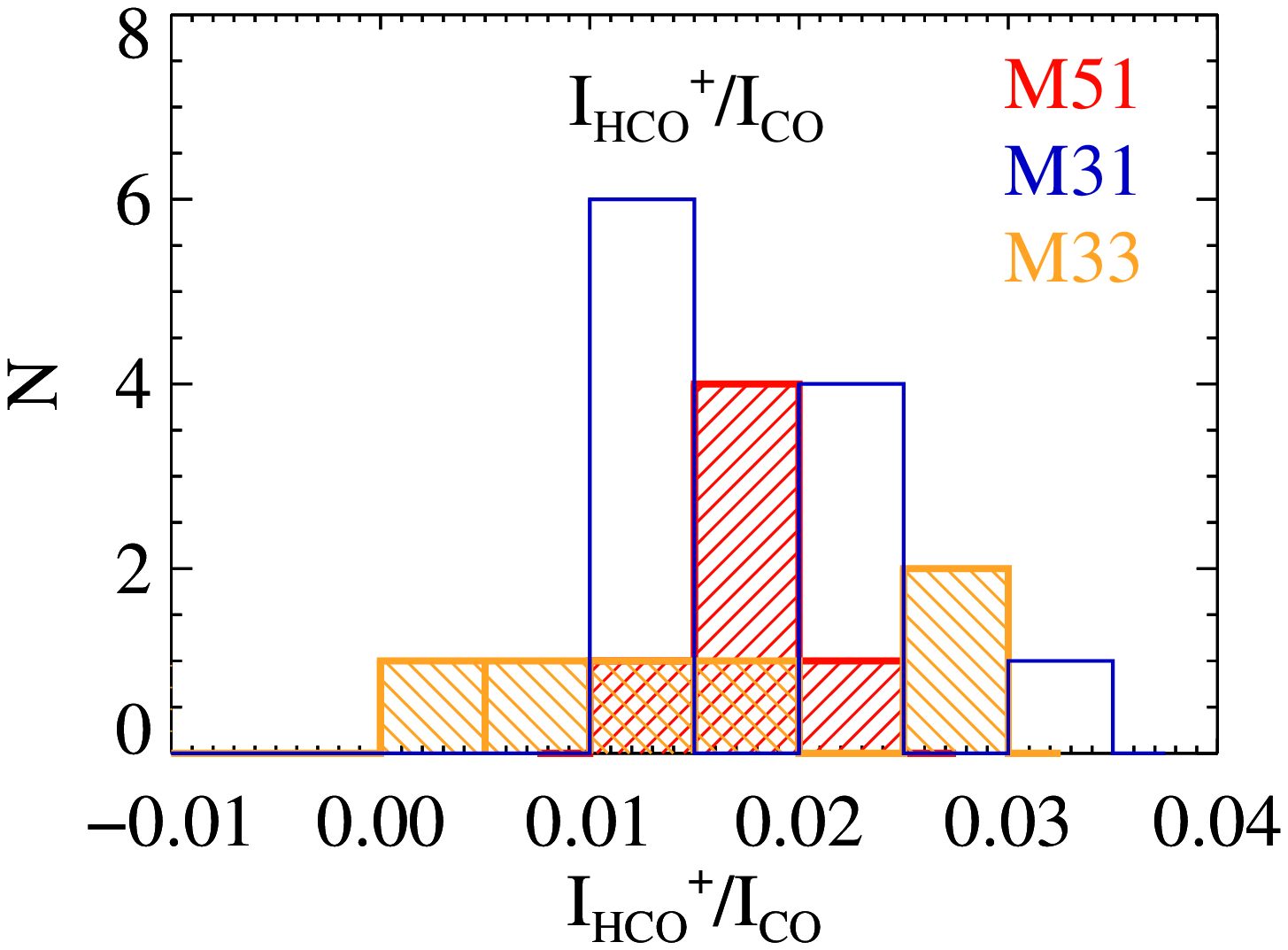}
\centering\includegraphics[angle=0,scale=.35]{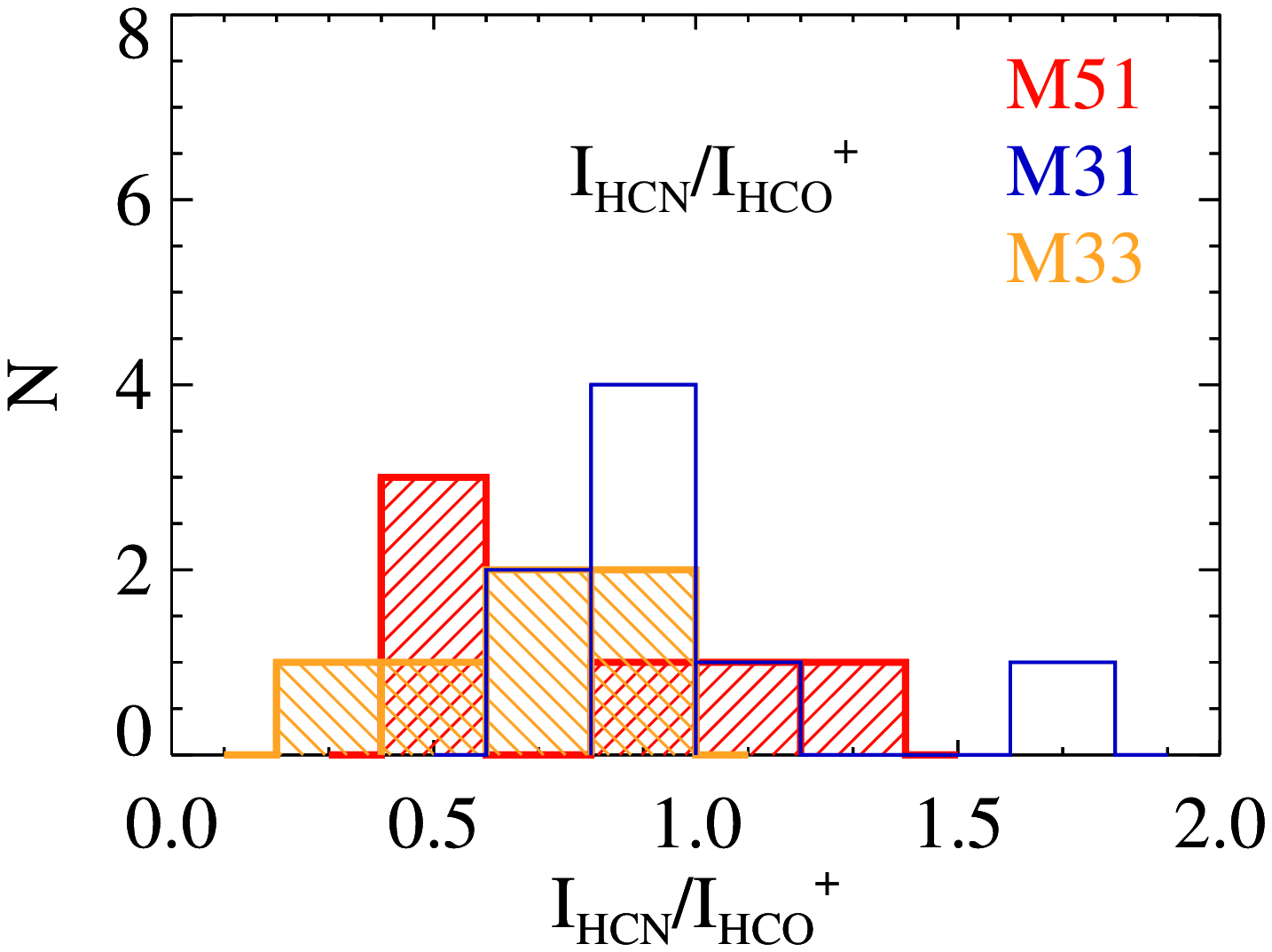}
\caption{HCN and HCO$^+$ integrated intensity distributions and line ratio ($I_{\rm CO}/I_{\rm HCN}$, $I_{\rm CO}/I_{\rm HCO^+}$ 
and $I_{\rm HCN}/I_{\rm HCO^+}$) 
distributions of the GMAs in M51 (red), M31 (blue) and M33 (yellow).}
\end{figure}

\begin{figure*}
\centering\includegraphics[angle=0,scale=.5]{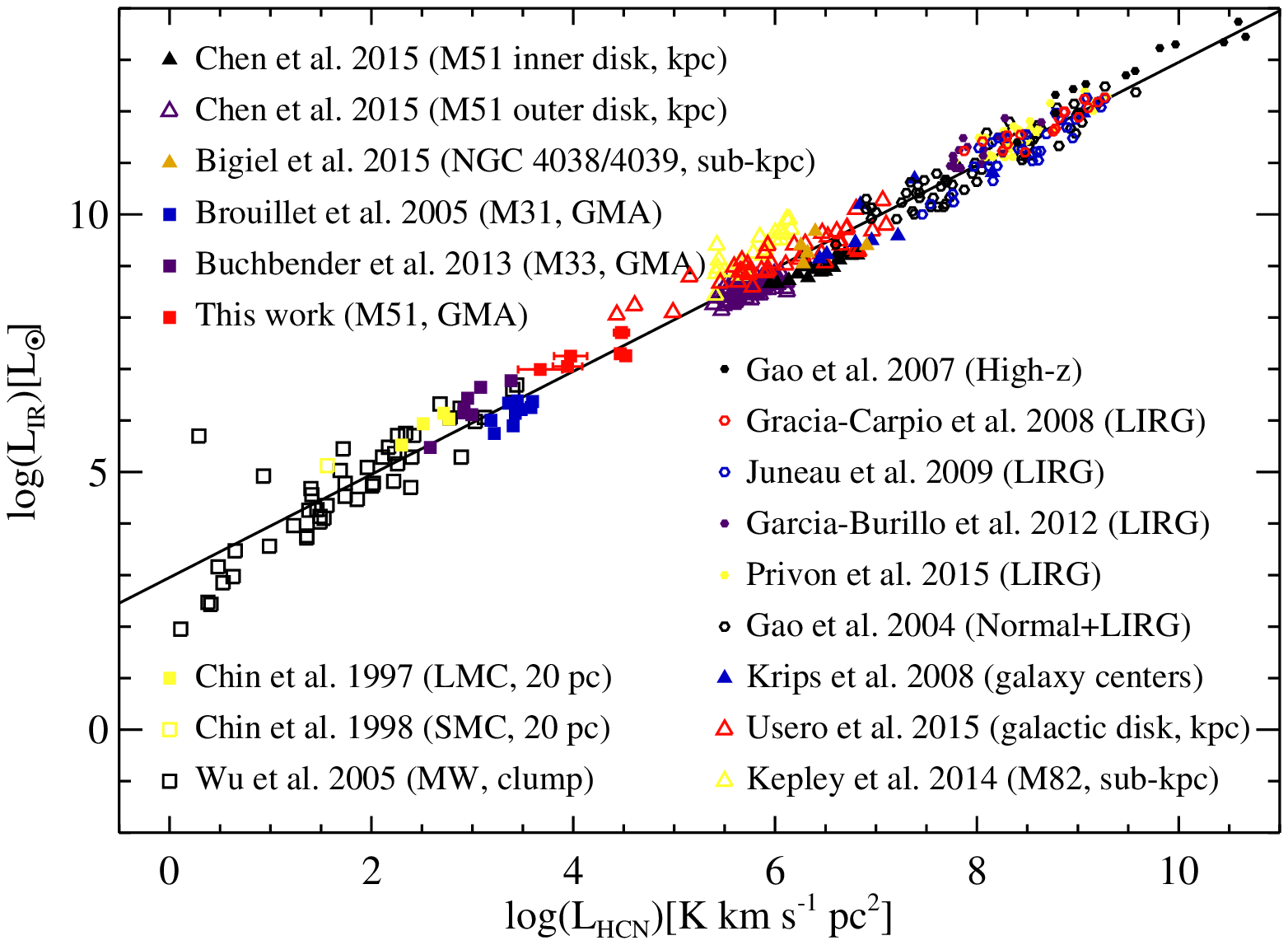}
\centering\includegraphics[angle=0,scale=.5]{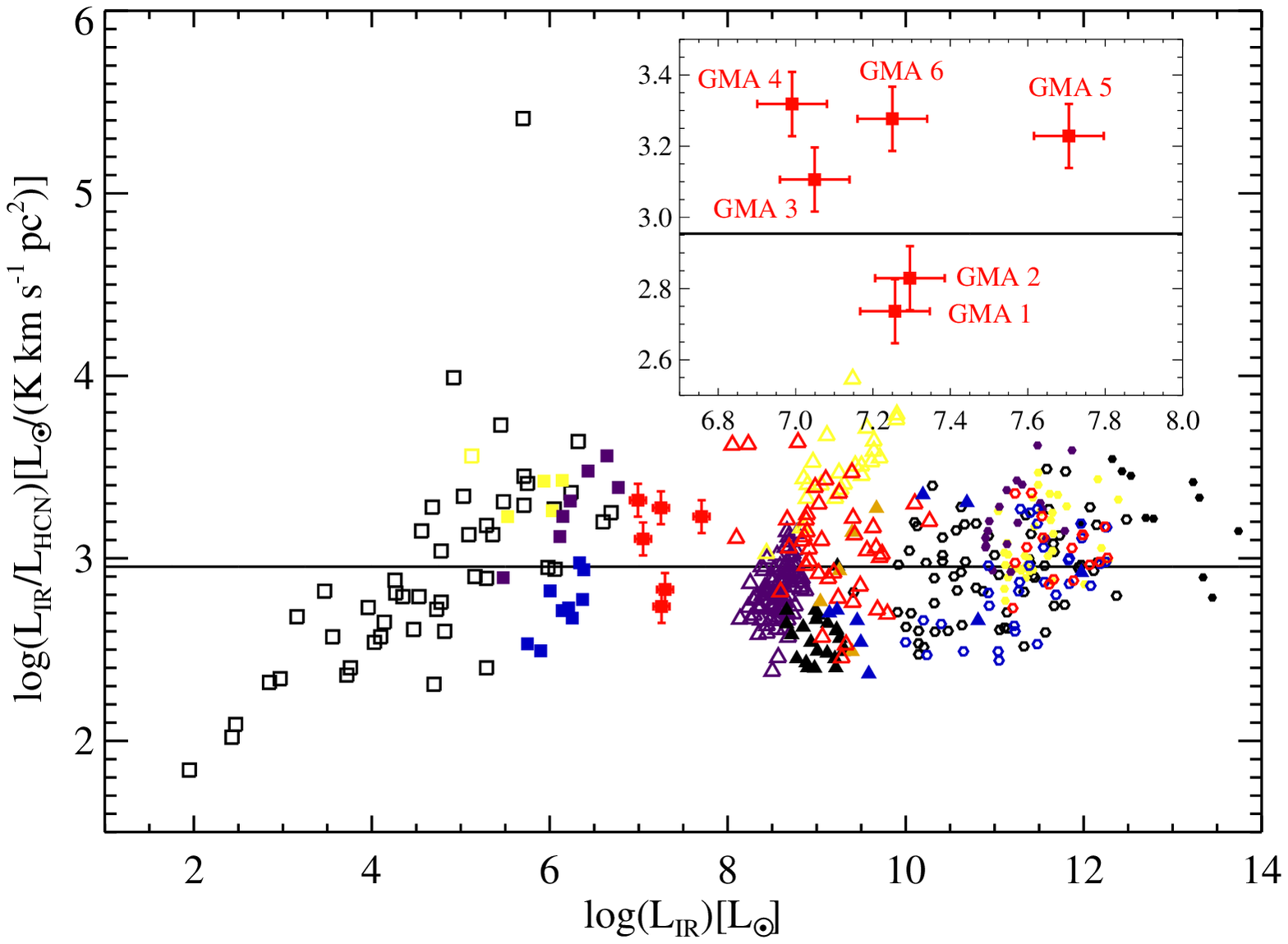}
\centering\includegraphics[angle=0,scale=.5]{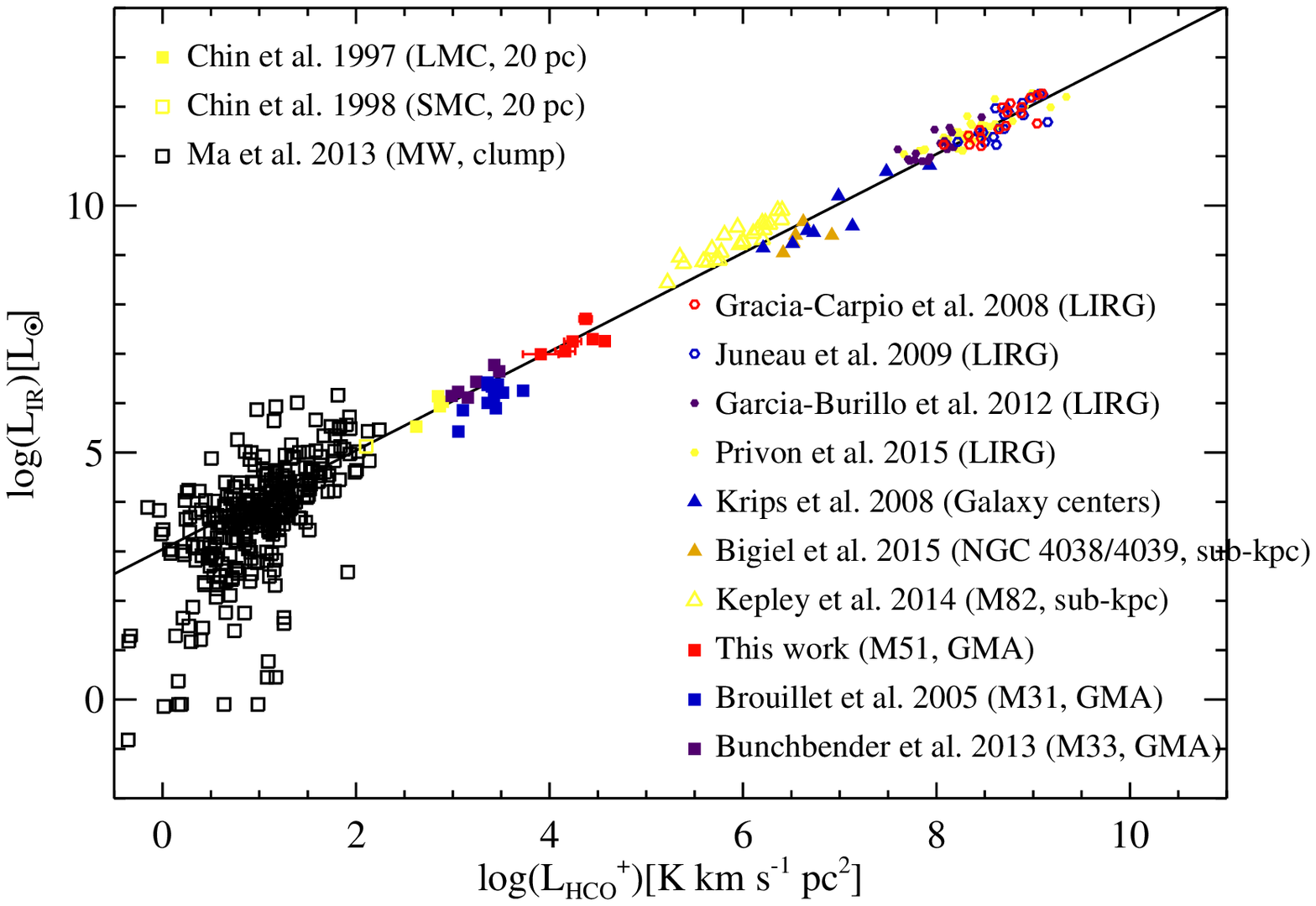}
\centering\includegraphics[angle=0,scale=.5]{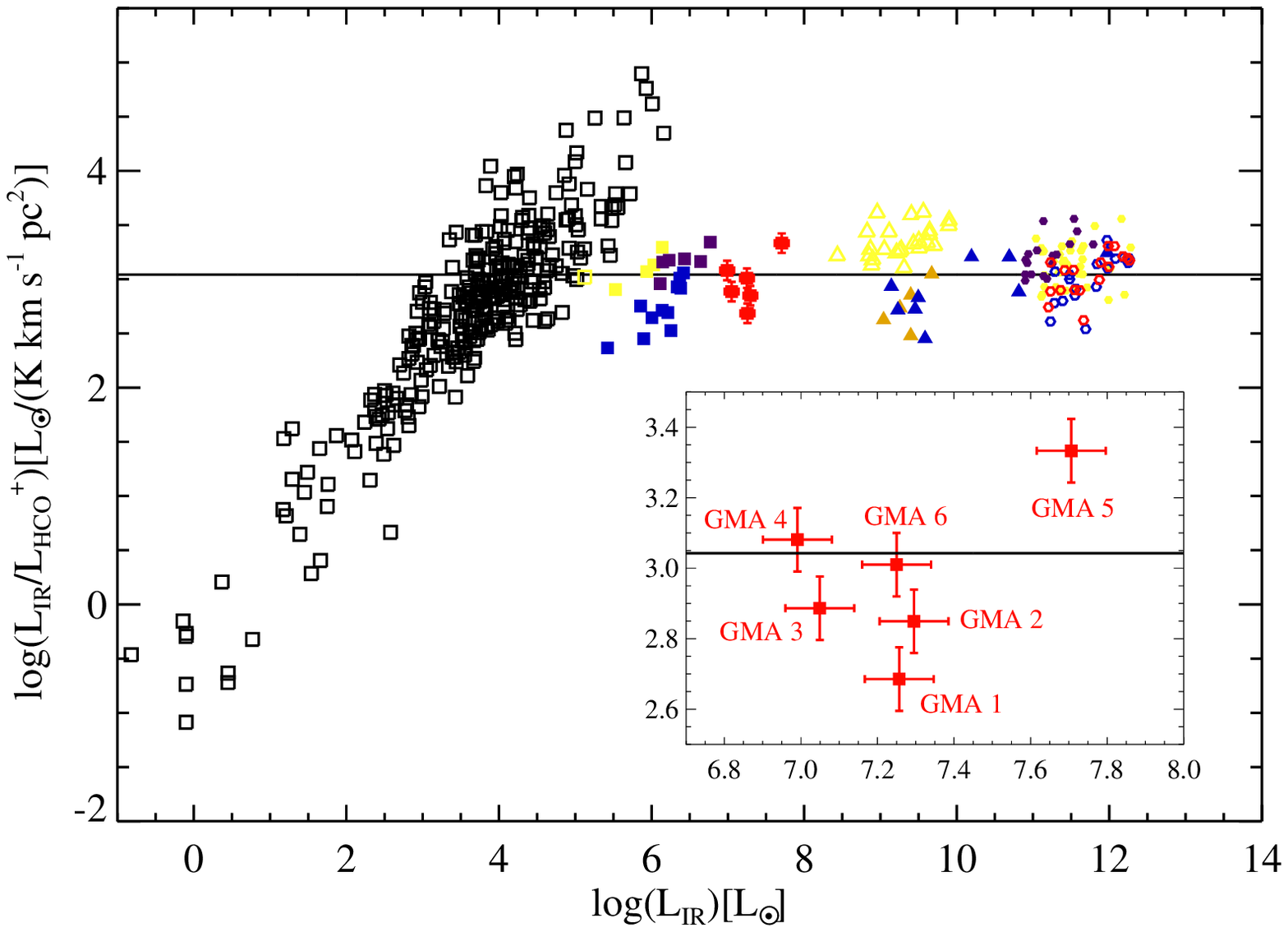}
\caption{(upper left) The correlation between $L_{\rm IR}$, tracing star formation rate, and $L_{\rm HCN}$, tracing dense gas mass, for clumps in the Milky Way \citep{Wu2010ApJS..188..313W}, giant molecular clouds in Magellanic Cloud \citep{Chin1997A&A...317..548C, Chin1998A&A...330..901C}, GMAs in nearby galaxies \citep[][and this work]{Brouillet2005A&A...429..153B, Buchbender2013A&A...549A..17B}, nearby galactic disk \citep{Kepley2014ApJ...780L..13K, Usero2015AJ....150..115U, Chen2015ApJ...810..140C, Bigiel2015ApJ...815..103B} and whole galaxies \citep{Gao2004ApJ...606..271G, Gao2007ApJ...660L..93G, krips2008ApJ...677..262K, Gracia-Carpio2008A&A...479..703G, Juneau2009ApJ...707.1217J, Garcia-Burillo2012A&A...539A...8G, Privon2015ApJ...814...39P}.
(lower left) The correlation between $L_{\rm IR}$, tracing star formation rate, and $L_{\rm HCO^+}$, tracing dense gas mass,  for clumps in the Milky Way \citep{Ma2013ApJ...779...79M},  giant molecular clouds in Magellanic Clouds \citep{Chin1997A&A...317..548C, Chin1998A&A...330..901C}, GMAs in nearby galaxies \citep[][and this work]{Brouillet2005A&A...429..153B, Buchbender2013A&A...549A..17B}, nearby galactic disk \citep{Kepley2014ApJ...780L..13K, Bigiel2015ApJ...815..103B} and whole galaxies \citep{krips2008ApJ...677..262K, Gracia-Carpio2008A&A...479..703G, Juneau2009ApJ...707.1217J, Garcia-Burillo2012A&A...539A...8G, Privon2015ApJ...814...39P}.
(right panel) The variations of $L_{\rm IR}$/$L_{\rm HCN}$ and $L_{\rm IR}$/$L_{\rm HCO^+}$, tracing star formation efficiency of dense gas, along with $L_{\rm IR}$, tracing the star formation rate, are shown for the same data in the left panels.
The insets show the zoom in plots of the data in this work. 
Only the uncertainties of HCN and HCO$^+$ for GMAs in M51 are shown in the plots.
Solid lines show the averaged ratios of IR/HCN and IR/HCO$^+$ for the \citet{Gao2004ApJ...606..271G} and \citet{Gracia-Carpio2008A&A...479..703G} samples.}
\end{figure*}

\subsection{HCN and HCO$^+$ Brightness and Line Ratios}

M31 \citep{Brouillet2005A&A...429..153B} and M33 \citep{Buchbender2013A&A...549A..17B} have been observed at virtually the same linear resolution as our M51 observations.  However, comparison of the M31, M33 and M51 disk GMAs in Figure 4 shows, the integrated intensities of the M51 GMAs are much higher in the HCN and HCO$^+$ lines. 

The dense gas fractions, as traced by  $I_{\rm HCN}/I_{\rm CO}$ and $I_{\rm HCO^+}/I_{\rm CO}$ , are compared in the lower part of Figure 4.  Interestingly, the $I_{\rm HCN}/I_{\rm CO}$ and $I_{\rm HCO^+}/I_{\rm CO}$ ratios are similar for the three galactic disks.
In M51, $I_{\rm HCN}/I_{\rm CO}$ ratios vary from 0. 007 to 0.021, which is similar to the ratio of Galactic disk \citep[$\sim$ 0.026,][]{Helfer 1997ApJ...478..233H}.

With the spectra in Fig. 3 and data in Table 2, we can estimate the CO-based total molecular gas mass (using a galactic conversion factor of N$_{H_2}$/$I_{co}$ = $2\times 10^{20} cm^{-2}/(K\ km/s)$) and the HCN-based dense gas mass \citep[using Eq. 8 from][]{Gao2004ApJS..152...63G}.
With these numbers, 
we obtain dense gas mass fractions from 2 to 7\% for the GMAs in M51. Thus, these observations suggest that a few percent of the mass in a GMC (or GMA, presumably a collection of neighboring GMCs) is in the form of dense gas.  
The simulations by \citet{Kroupa2001MNRAS.321..699K} find that
30\% of dense gas mass is turned into stars, such that 0.6--2.1\% of the total cloud mass (2--7\% $\times$ 0.3) form stars in our sample, which is in reasonable agreement with the estimate of \citet{Zuckerman1974ApJ...192L.149Z} that about 1\% of the mass of a molecular cloud is converted into stars.

The $I_{\rm HCN}/I_{\rm HCO^+}$ line ratios are also compared for the GMAs in M51, M31 and M33. 
All but two of the GMAs  show similar $I_{\rm HCN}/I_{\rm HCO^+}$ ratios, ranging from 0.4 to 1.0. 
The two GMAs in M51 and M31 hosting powerful H\uppercase\expandafter{\romannumeral2} regions (star forming regions) show high $I_{\rm HCN}/I_{\rm HCO^+}$ ratios (1.21 and 1.72).

\subsection{Star Formation Rate vs. Dense Gas Mass}

To compare the dense gas mass with the SFR of the GMAs in the outer spiral arm of M51, we use the HCN and HCO$^+$ luminosities to trace the dense gas mass and IR luminosity to trace the SFR. The line luminosities are calculated as $L[K\ km\ s^{-1}\ pc^2]= 23.5\Omega [arcsec^2]d_L^2[Mpc]I[K\ km\ s^{-1}]$, where $\Omega$ is the solid angle of the GMA, $d_L$ is the luminosity distance of M51 \citep[7.6 Mpc,][]{Ciardullo2002ApJ...577...31C} and $I$ is the integrated intensity calculated as in Table 2. The line luminosity uncertainties are determined from the same formula, substituting $I_{rms}$ for $I$ and taking $\Omega$ as either the solid angle of GMA or the $5.6\arcsec$ beam size, whichever is larger. 
The IR luminosities are derived from Herschel 70 $\micron$ data following the conversion function shown in Table 2 of \citet{Galametz2013MNRAS.431.1956G}.
The uncertainty in $L_{\rm IR}$ comes from the scatter in the  conversion of 70 $\mu$m to IR \citep[0.09 dex,][]{Galametz2013MNRAS.431.1956G}. This uncertainty dominates the measurement error at 70 $\mu$m (0.01 to 0.05 dex).
When calculating luminosities, all the maps have been smoothed to the angular resolution of the 70 $\mu$m data (5.6$\arcsec$) and the results are listed in Table 3.

Our observations of the 6 GMAs fill the gap in IR-HCN and IR-HCO$^+$ relations  between the large-scale observations, kpc or larger, and the Galactic measurements (see Figure 5).
Both the L$_{\rm IR}$--L$_{\rm HCN}$ and L$_{\rm IR}$--L$_{\rm HCO^+}$ relations in the outer disk GMAs are consistent with the proportionality between L$_{\rm IR}$ and dense gas mass established globally in galaxies.
The observations of the GMAs presented here are quite close to the average IR-HCN relation with no obvious shift or trend, unlike some of the Galactic observations or observations of galactic centers which tend to have low IR/HCN flux ratios.  

There is still some scatter in IR/HCN and  IR/HCO$^+$  ratios in the GMAs of the outer disk as shown by the insets in Figure 5.
Both IR/HCN and IR/HCO$^+$ ratios of GMA 5 are 3 times higher than GMA 1 and 2. This could be because massive stars are forming in GMA 5 as it hosts an H\uppercase\expandafter{\romannumeral2} region. The IR/HCN ratio of GMA 3, 4 and 6 is almost the same as GMA 5, but their IR/HCO$^+$ ratio is 1/2 of GMA 5. It is consistent with the lower HCN/HCO$^+$ ratio of GMA 3, 4 and 6  (about 0.5) than GMA 1, 2 and 5 (about 1.0).
The results do not change when we take extinction-corrected H$\alpha$ \citep{Calzetti2007ApJ...666..870C} or FUV \citep{Liu2011ApJ...735...63L}  instead of IR to trace SFR.

\begin{deluxetable*}{lllllll}
\tabletypesize{\scriptsize}
\tablecaption{CO, HCN, HCO$^+$, HNC intensities}
\tablewidth{0pt}
\tablehead{
\colhead{} & \colhead{GMA 1} &  \colhead{GMA 2} & \colhead{GMA 3}  & \colhead{GMA 4} & \colhead{GMA 5}&  \colhead{GMA 6}
}
\startdata
\colhead{$I_{\rm CO}$ [K\ km/s]} & $46.2\pm2.8$ & $47.8\pm4.0$ & $35.0\pm3.9$ & $28.3\pm4.0$ & $30.1\pm2.7$ & $34.0\pm3.2$  \\
\colhead{$I_{\rm HCN}$ [K\ km/s]} & $0.53\pm0.06$ & $0.97\pm0.07$ & $0.30\pm0.09$ & $0.25\pm0.10$ & $0.64\pm0.12$ & $0.29\pm0.12$  \\
\colhead{$I_{\rm HCO^+}$ [K\ km/s]} & $0.63\pm0.06$ & $0.88\pm0.09$ & $0.60\pm0.11$  & $0.60\pm0.11$ & $0.53\pm0.08$ & $0.60\pm0.10$ \\
\colhead{$I_{\rm HNC}$ [K\ km/s]} & $0.21\pm0.04$  & $0.42\pm0.06$ & $0.10\pm0.07$ & $0.13\pm0.07$ & $0.05\pm0.07$  & $0.06\pm0.09$ \\
\colhead{$I_{\rm HCN}/I_{\rm CO}$} & $0.011\pm0.002$ & $0.020\pm0.003$ & $0.009\pm0.004$ & $0.009\pm0.005$ & $0.021\pm0.006$ & $0.009\pm0.004$  \\
\colhead{$I_{\rm HCO^+}/I_{\rm CO}$} & $0.014\pm0.002$ & $0.018\pm0.003$ & $0.017\pm0.005$ & $0.021\pm0.007$ & $0.018\pm0.004$ & $0.018\pm0.005$  \\
\colhead{$I_{\rm HCN}/I_{\rm HCO^+}$} & $0.84\pm0.18$ & $1.10\pm0.19$ & $0.50\pm0.24$ & $0.42\pm0.24$ & $1.21\pm0.41$ & $0.48\pm0.28$ 
\enddata
\tablecomments{
In calculating the intensities, the CO map has been smoothed to $4.9\arcsec\times3.6\arcsec$ to match the angular resolution of HCN, HCO$^+$ and HNC maps.
The integrated intensities were measured as $I = \int Td$V over the emission velocity range. 
The uncertainties are calculated as $\delta =  T_{\rm rms}\sqrt{W\delta_c}$.  The rms temperature uncertainty ($T_{\rm rms}$) is calculated from the non-emission channels of the spectra. W and  $\delta_c$ are the line width and channel width in velocity.
}
\end{deluxetable*}

\begin{deluxetable*}{lllllll}
\tabletypesize{\scriptsize}
\tablecaption{CO, HCN, HCO$^+$ and IR luminosities}
\tablewidth{0pt}
\tablehead{
\colhead{} & \colhead{GMA 1} &  \colhead{GMA 2} & \colhead{GMA 3}  & \colhead{GMA 4} & \colhead{GMA 5}&  \colhead{GMA 6}
}
\startdata
\colhead{$\Omega[arcsec^2]$} & $47.16$ & $26.28$ & $20.88$ & $11.52$ & $37.80$  & $28.44$ \\
\colhead{$L_{\rm CO\ \ }$ [$10^5$ K\ km/s\ pc$^2$]} & $26.9\pm1.7$ & $15.4\pm1.7$ & $8.6\pm1.6$ & $3.6\pm1.8$ & $13.3\pm1.2$ & $11.1\pm1.4$  \\
\colhead{$L_{\rm HCN\ }$ [$10^3$ K\ km/s\ pc$^2$]} &$33.2\pm3.1$ &$29.3\pm2.8$ & $8.8\pm3.5$ & $4.7\pm3.1$ & $30.0\pm5.6$  & $9.4\pm4.3$   \\
\colhead{$L_{\rm HCO^+}$ [$10^3$ K\ km/s\ pc$^2$]} & $37.3\pm2.8$ & $27.9\pm3.3$ & $14.5\pm3.9$  & $8.1\pm4.3$ & $23.6\pm3.7$ & $17.4\pm3.9$ \\
\colhead{$log(L_{\rm IR})$ [$L_\odot$]} & $7.26\pm0.09$ & $7.30\pm0.09$ & $7.05\pm0.09$ & $6.99\pm0.09$ & $7.71\pm0.09$  & $7.25\pm0.09$ \\
\colhead{$log(L_{\rm IR}/L_{\rm HCN})$ [$L_\odot/(K\ km/s\ pc^2)$]} & $2.74\pm0.09$ & $2.83\pm0.09$ & $3.11\pm0.09$ & $3.32\pm0.09$ & $3.23\pm0.09$  & $3.28\pm0.09$ \\
\colhead{$log(L_{\rm IR}/L_{\rm HCO^+})$ [$L_\odot/(K\ km/s\ pc^2)$]} & $2.69\pm0.09$ & $2.85\pm0.09$ & $2.89\pm0.09$ & $3.08\pm0.09$ & $3.33\pm0.09$  & $3.01\pm0.09$
\enddata
\tablecomments{ 
When calculating luminosities, all maps have been smoothed to the angular resolution of 70 $\mu$m (5.6$\arcsec$). 
The line luminosities are calculated with the function of $L[K\ km\ s^{-1}\ pc^2]= 23.5\Omega [arcsec^2]d_L^2[Mpc]I[K\ km\ s^{-1}]$, where $\Omega$ is the solid angle of the GMA, $d_L$ is the luminosity distance of M51 \citep[7.6 Mpc,][]{Ciardullo2002ApJ...577...31C} and $I$ is the integrated intensity calculated with the same way as Table 2.  
The uncertainty in $L_{\rm IR}$ comes from the scatter in the  conversion of 70 $\mu$m to IR \citep[0.09 dex, ][]{Galametz2013MNRAS.431.1956G}. This uncertainty dominates the measurement errors in 70$\mu$m (0.01 to 0.05 dex).
}
\end{deluxetable*}

Figure 5 shows not only the extragalactic data points sampling objects with IR luminosities above $10^5$L$_\odot$ at a variety of scales, but also many Galactic observations, enabling sampling down to very low luminosities.
While some of the apparent scatter in Figure 5 could simply be due to noise, Wu et al. (2005) and Ma et al. (2013) show that the IR luminosity at the weak end appears to be lower than the prediction from the HCN (or HCO$^+$) luminosity by the IR--HCN (or IR--HCO$^+$) linear relation.
At very small scales or low luminosities, this non-linearity could be introduced by the incomplete sampling of the stellar IMF \citep[e.g.][]{Kennicutt2012ARA&A..50..531K} and thus poor measurement of the SFR which depends strongly on the high stellar mass sampling. 
\citet{Kennicutt2012ARA&A..50..531K} suggested that the SFR should be larger than $0.001\sim0.01\ M_\odot\ year^{-1}$ to completely sample the stellar IMF. The SFR of our GMAs are about $0.002\sim0.01\ M_\odot\ year^{-1}$ \citep[derived from IR with Eq. 9 in][although mainly adopted/used globally in star-forming galaxies]{Gao2004ApJ...606..271G}, so it is not clear whether the scatter in IR/HCN and IR/HCO$^+$ ratios in these GMAs could be due to incomplete sampling.

\section{ SUMMARY }

We mapped a selected region on the outer spiral arm of M51 in HCN(1--0), HCO$^+$(1--0) and HNC(1--0) using the NOEMA interferometer with an angular resolution of 4\arcsec ($\sim 150$ pc). \\(1) We detected bright emission of HCN and HCO$^+$ in 6 GMAs defined by CO(1--0) data, while HNC emission is only detected in the two brightest GMAs. 
\\(2) The HCO$^+$ spatial distribution is generally broader than that of HCN and HNC.
One of the GMAs hosts a powerful H\uppercase\expandafter{\romannumeral2} region and HCN is stronger than HCO$^+$ there.
\\(3) The GMAs in M51 are brighter in both HCN and HCO$^+$ than the GMAs in M31 and M33, but the ratios of CO/HCN, CO/HCO$^+$ and HCN/HCO$^+$ are similar for the three objects. 
\\(4) Combined with Herschel 70$\mu$m data, we find that both the L$_{\rm IR}$--L$_{\rm HCN}$ and L$_{\rm IR}$--L$_{\rm HCO^+}$ relations in GMAs of M51 follow the proportionality between the L$_{\rm IR}$ and the dense gas mass established globally in galaxies within the scatter. \\
(5) The IR/HCN and  IR/HCO$^+$  ratios of the GMAs vary by a factor of 3, probably depending on whether massive stars are forming or not. 

\acknowledgments

We appreciate the generous support from IRAM staff and GILDAS team during the observations and data reduction. This work is supported by the program for Outstanding PhD candidate of Nanjing University, the National Natural Science Foundation of China (grants 11173059, 11390373, 11420101002, 11273015 and 11133001), CAS pilot-b program (No. XDB09000000) and the National Basic Research Program (973 program No. 2013CB834905).
We are very grateful to Campus France for Xu Guangqi grant 34454YG which helped fund the travel necessary for this work.

\clearpage
\end{document}